\documentclass[12pt]{report}
\usepackage{amssymb, amsfonts, amscd}
\begin{document}
\title{A Dynamical Similarity Approach to the Foundations of Complexity and Coordination in Multiscale Systems\footnote{A University Scholar Project submitted in partial fulfillment of the requirements of the Bachelor of Arts (Mathematics) degree as University Scholar at University of Connecticut in August 2003.}}
\author{Abhijnan Rej\\ Department of Mathematics\\ University of Connecticut\\ Storrs, CT 06269--3009\\
\\ rej@math.uconn.edu }
\date{}
\maketitle

\tableofcontents

\chapter{Introduction}
\section{Order and its generation}
The study of purposive activity of the central nervous system can be reduced to the search for answers to two complimentary problems. The first problem is that of \emph{coordination} which is, roughly, the following: inorder to satisfy everyday demands, an animal routinely ``brings together'' segments of its body with one another and with the surrounding layout of surfaces (Turvey, 1990) The resultant movements (when they are thus generated) are ordered in space and time, even though they arise due to very many microscopic ensembles (neurons, neuromuscular elements, tissues, \ldots) acting within, and upon, one another. The second problem is the problem of \emph{control}: it is the problem of regulation of purposive (i.e., voluntary) action such that it satisfies the constraints imposed by a given goal. When elaborated, the problem of control includes all actions by any arbitrary (natural or artificial) system that is goal-directed. This is, then, the problem of \emph{cybernetics}. One can, therefore, see the problem of pruposive activity reducing to the study of order (coordination) and its generation (control) in \emph{sufficiently complex systems}.
\section{The control paradigm}
The two complementary problems have led to two disparate groups of theories and approaches. Approaches to the problem of control are, historically, older than the approaches to the problem of coordination (at least in the modern sense of the term.) The simplest approach to the problem of control is evoking some higher (structural) unit of the central nervous system, the \emph{central pattern generator}, (for example, in the cortex) which generates commands for  the execution of purposive action. These commands are then transmitted to the organ that is to physically ``carry out'' the action (e.g., limbs) via the remaining neural mechanisms whose roles are precisely to do the same (Gelfand et.al., 1971). This view of control stems from the classical work of Fritsch and Hitzig and manifests itself fully in the work of Sherrington, Pavlov and others. In more modern terminology, one sees this central pattern generator as being a \emph{motor program} (Keele, 1968.) A motor program, like a computer program, is often seen as a symbol string where ``the symbols are commands that, when executed by the neuromuscular apparatus, dictate particular patterns of muscular activity that \ldots produce a desired movement'' (Riley \& Turvey, 2001.)
\section{The coordination paradigm}
Approaches to the problem of coordination are more recent, and involve viewing systems that exhibit coordination (the so-called \emph{coordinative structures}) as \emph{dissipative structures} that exist under far-from-equilibrium conditions (Kugler et.al, 1979.) Consequently, under this philosophy, coordinative structures are modelled using synergetics-- typically, an order parameter is identified that satifies a nonautonomous nonlinear (partial) differential equation.The basic idea in such approaches is to identify an order parameter that somehow \emph{enfolds in itself} the contributions of layers of physiological structures (neurons, neuromuscular tissues, \ldots). There is no systematic way to identify the order parameter that maximally encapsulates ``useful'' information about the processes that are to be modelled. The canonical example of this approach is the Haken-Kelso-Bunz model of interlimb coordination.
\section{Non-reductionism and touchstones for  multiscale coordination}
The Haken-Kelso-Bunz model approaches the problem of interlimb coordination at the level of end-effectors. The order paramter $\phi$ is the relative phase of oscillations of the limbs and satifies
\begin{equation}
\frac{d\phi}{dt} = -\frac{dV}{d\phi}
\end{equation}
where $V(\phi)$ is the potential function
\begin{equation}
V(\phi) = -A \cos \phi -B \cos 2\phi
\end{equation}
$A$ and $B$ are free parameters (see Haken, Kelso \& Bunz, 1984). By the very nature of this model, one does not gain much information about the underlying neural processes that coallesce to produce behavior at the level of the end-effectors (here, the limbs). In order to address this issue and make the HKB model congruous with some purported dynamical model at the neural level, Beek, Peper and Daffertshofer (Beek, Peper \& Daffertshofer, 2002) have proposed a two-tiered organization of rhythmic interlimb coordination that couples dynamical behavior at the level of some (unspecified) neural substrate to the classical HKB model.Their rationale for adding this new level of "neural oscillators" was essentially experimental. The HKB model, as it stands, fails to account for   a number of experimental observations such as the existence of non-monotonic amplitude-frequency relations (Kay et.al., 1987) and (phase-dependent) phase shift after perturbation (Kay et.al., 1991).
\par
There are two main reasons why one must attempt to build models that are multi- and cross- scale.
\begin{itemize}
\item{\emph{Stochasticity} Motor control, in general, exhibits piece-wise determinism. In particular, studies have revealed the essentially stochastic nature of fluctuations of the center of pressure during quiet standing (Riley \& Turvey, 2001).  Very often, the origin of disorder comes from the observer-theorist's "deliberate refusal to specify (and follow) the locations and behavior of particles on the microlevel that surround a given particle and interact with it" (Mineev-Weinstein, 1996). Therefore, any complete theory of biological coordination that is (at least partly) stochastic must take into account the multitude of degrees of freedom at a microscopic level.  These microscopic degrees of freedom may be suppressed in the macroscopic analysis; one can conjecture that 'noise' in systems arise due to the so-called "coarse-graining" of the system (Tomita, 1984). Noise generally manifests itself as Langevin terms as in the Fokker-Planck equations of Schoner, Haken and Kelso (1986). }
\item{\emph{Evolutionary holism} The most important reason, however, behind looking for a multiscale coordination dynamic is the following: "large brains gain their powers not so much by the number of neurons they contain, but by the number of scales of neural organization they support" (Alexander \& Globus, 1996). If one believes that evolution has spent an apparently inordinately large amount of time trying to solve the problem of locomotion (Brooks, 1991), then an analysis of interlimb coordination (instrumental in solving the problem of gait) that is sensitive to evolutionary history must try to include the various scales of organization that may have given us the kind of coordination we observe. The search of the laws of multiscale coordination, thus, is also in line with Edward Wilson's principle of \emph{evolutionary holism}: "\ldots complex wholes will usually be developed incrementally over evolutionary time, and that the various intermediate forms must themselves be whole, robust systems \ldots" (Clark, 1999).}
\end{itemize}
However, oscillator models that couple scales may fall prey to a version of morphological reductionism: That is, some (micro)scale of the model is seen as privileged or fundamental (often some uncharacterized ``canonical ensemble'' cf. Frank, Daffertshoffer and Beek, 2000) and the remaining scales are seen to be related in a linear causal fashion to that scale. Additionally, they tie in various physiological scales of a coordinative agent by introducing successive layers of oscillators whose introduction have no \emph{\'{a} prior\'{i}} legitimization.
\par
A theory of multiscale coordination that is to avoid this trap of morphological reductionism must satisfy two touchstones:
\begin{itemize}
\item{The theory must address whether the system/subsystem distinction in its models are an ontological dichotomy or an epistemological neccessity, and incorporate the same in its modelling strategy. This is because of the following reason: morphological reductionism, by definition, asserts the existence of elemental structural units. These structures are seen as primary causal agents. One often identifies these primary structural units of a given system as its subsystems. The abhorrence of primary structures, then, neccessarily entail the banishment of the system/subsystem dichotomy.}
\item{The theory must incorporate a minimally small set of maximally entailed free parameters in its models. In multiscale coordination, the free parameters at each scale must be entailed by free parameters at other scales.}
\end{itemize}
In this Project, we identify a possible candidate for a non-reductionistic theory of multiscale coordination. The proposed theory obeys the principle of dynamical similarity-- the equations of motion for different scales of physiological organization are "similiar". This similarity is formalized through the notion of \emph{topological conjugacy}, and following Rosen (1978), we say that the proposed theory obeys the \emph{principle of dynamical similarity}.

\chapter{Defining Complexity}
\section{Complex systems?}
At the heart of any discussion of how a given mutiscale system with many components coordinate and control the observed phenomenology lies the question of the \emph{complexity} of that system. This is mainly because of a cultural 'paradigm-shift' in the natural sciences where any system that is structurally constituted of very many interacting parts and with cross-scale interactions is seen as being outside the purview of classical simple systems that generally mimic the composition of simple Newtonian systems. Indeed, with the introduction of ideas from nonlinear dynamics to biology, it has become quite fashionable to assert that the richness of the behavior of biological systems has something to do with its complexity (cf. Goodwin \& Sole, 2000.) Yet, the very definition of the word 'complexity' is hard to come by. In this Chapter, I wish to discuss two approaches to that problem of complexity, one of \emph{computational mechanics} (Crutchfield, 1994) and \emph{computational complexity of natural systems} (Wolfram, 2002 and Chaitin, 1987), collectively termed as the computational approach, and the other of \emph{impredicative complexity} (Rosen 1978, 2001a, 2001b). I will examine the importance of the two approaches in creating a viable and non-reductionist theory of multiscale coordination.
\par
The idea that complexity somehow ``lies in the eye of the beholder'' is a natural and intuitive notion, one that stems from our everyday usage of the term. Afterall, we marvel at a chess Grandmaster for \emph{easily} solving a (chess) puzzle that seems immensely complex to us. For God, every phenomena, no matter how complex, will appear simple-- this has been the leading credo of the classical mechanical view of the Universe as it was seen in the seventeenth and eighteenth centuries; recall Laplace saying that with Newton's equations of motion, one can view the entire Universe with all its details, as merely being a clock.
\par
The idea that most nonlinear dynamical systems are sensitively-dependent on the initial data that is ``given'' to them has effectively ended the Laplacian program in science. We now know that even very simple one-dimensional maps (like the quadratic map studied by Robert May in the early 1970s) can produce (when the free-parameters in those maps are ``tweeked'') a extremely rich spectra of behavior that ranges from very simple and regular (limit-cycles) to completely random (chaotic).
\par
However, when we try to summarize and describe the behavior that is 'sandwiched' between these two extremes, we are often at a loss. Typically, this ``intermediate'' part of the spectra exhibits both order and disorder, both the presence and absence of structure. To qualify this claim, let us look into Wolfram's classification scheme for the set of all elementary cellular automata (ECA). Let us assume that we have run each of the all possible ECA rules (256 rules) for $n$ steps of evolution.
\begin{itemize}
\item{Class I = all ECAs that die out.}
\item{Class II = all ECAs that start producing regular and/or self-similiar behavior that stabilizes and ``settle''.}
\item{Class III = all ECAs that exhibit ``islands'' of structures that are localized, yet these structures are neither homogeneously nor isotropically distributed.}
\item{Class IV = all ECAs where we cannot detect any structure whatsoever, i.e., ECAs that are random.}
\end{itemize}
I phrase the Wolfram classification (Wolfram, 1994) in this way for two reasons: (a) this is how the classification was first described, as a sort of a ``visual reaction'', and (b) the very way the classification is described suggests the central theme of this Chapter: will observer X classify a given ECA in the same way (i.e., put it into the same class) as any other observer Y? In other words, \emph{is there a universal and quantitative characteristic of any given ECA that is invariant for all observers and allow any two of them to use this charateristic to classify that ECA into the same class?}
\par
What is being alluded to by the aforementioned question is the notion of \emph{subjectivity}\footnote{It seems pertinent to mention that the question of classification could be equally applied to the behavior of continous dynamical systems; indeed, one way to view continous systems is to view them as ECAs taken to a continuum limit. For example, one can obtain the Navier-Stokes equations by taking a suitable continuum limit of certain ECAs.}. In order to study this question further, I introduce a 'pictorial' notational system (\emph{the inference chain} and \emph{action chain}) that allows us to succinctly discuss the results of Crutchfield, 1994, Chaitin, 1987, Wolfram, 2002 and Rosen, 2001a, 2001b in a unified framework. My intention is to frame the notions of complexity, computation and measurement problems (both in quantum and classical-chaotical systems) in a single package. The principle advantage of doing so is to explore the duality of measurement as inference (Chaitin, 1987 and Wolfram 2002) on one hand and measurement as action (Crutchfield, 1994 and Rosen, 1978) on the other. 
\section{The inference chain}
Some notation: Let $\{O_i\}_{i\in A}$ be the set of observables, $A$ some index set. Let $\{M_j\}_{j\in B}$ be the set of meters, $B$ some other index set. By a meter, I mean a measurement device that inputs some phenomena and outputs some charateristic of that phenomena. 
\par
In essence a \emph{measurement} of $O_i$ is \emph{infering through $M_j$ certain properties of $O_i$}. Operationally, what a meter does (when applied to $O_i$ instantenously at a given region of the Universe) is it gives us a real number that we say is a measurement of $O_i$\footnote{The methodology of empirical natural sciences largely boil down to choosing suitable meters that provide us maximal information about $O_i$}. I denote this as:
$$
\begin{CD}
O_i @>\textrm{input}>> M_j @>\textrm{output}>> r (\in \mathbb{Q}).
\end{CD}
$$
Application of $M_j$ to $O_i$ $t$ times will give us a time series. We denote this by
$$
\begin{CD}
(O_i @>\textrm{input}>> M_j)^t @>\textrm{output}>> \{r_1, r_2, \ldots, r_t\} (:=\mathcal{T}).
\end{CD}
$$
I call chains like the ones above \emph{inference chains} because we use the meter to infer about the observable. Now, a natural question about inference chains is the following: given a time-series, what can we say about the observable? In other words, can we \emph{reconstruct} the dynamics that $O_i$ obeys on the basis of this time-series? The answer is yes. In the past decade, a number of methods have been developed that center around Taken's theorem, and allow us to reconstruct the geometry of the phase space of $O_i$ from time-series in a one-to-one sense (Packard et.al., 1980.) 
\par
This notation also allows us to abstractly capture the essence of \emph{phase transitions}. To see this consider $O_i$ to be a variable such that $r \in \{-1,1\}$. Thus the inference chain is going to be of the form
$$
\begin{CD}
(O_i @>\textrm{input}>> M_j)^t @>\textrm{output}>> \{-1,1\}^t
\end{CD}
$$
(${\{-1,1\}}^t$ means a $t$-tuple with entries either $-1$ or $+1$. If we obtain a time-series $\mathcal{T} = \{-1,1,-1,1,-1,1, \ldots, 1,1,1,1, \ldots\}$ we say that observable $O_i$ has undergone a phase transition.
\par
We now introduce another set of objects that I call \emph{recognition devices} each denoted by some $R_k$. These devices input $\mathcal{T}$ and sort the observable from which $\mathcal{T}$ is obtained into any of $n$-bins uniquely. We, at this point, do not think of these bins as being anything other than placeholders for observables. In the inference chain notation,
$$
\begin{CD}
(O_i @>\textrm{input}>> M_j)^t @>\textrm{output}>> \mathcal{T} @>\textrm{input}>> R_k @>\textrm{output}>> \textrm{some bin}.
\end{CD}
$$
\par
The natural question here is the following: how are we to construct $R_k$? One answer is provided by the notion of \emph{Kolmogorov-Chaitin complexity} or \emph{algorithmic complexity} in general. Chaitin's construction of $R_k$ will be viewing $R_k$ as the shortest algorithm (the program length in bits is the smallest) that inputs $\mathcal{T}$ with its length in bits $\alpha$ and outputs a description of $\mathcal{T}$ that is of length $\beta < \alpha$. The bin into which $\mathcal{T}$ would be put is, for Chaitin, uniquely specified by the length of the smallest algorithm that achieves this. The complexity of $\mathcal{T}$, for Chaitin, is  
$$ C_{\mathcal{T}} = \textrm{smallest } \sharp(R_k) $$
where $\sharp(R_k)$ is the length of $R_k$ such that the description of $\mathcal{T}$ (in bits) provided by inputing $\mathcal{T}$ into $R_k$ is shorter than $\mathcal{T}$ itself. 
\par
Now as Chaitin and others observed, suppose $\mathcal{T}$ is random in the sense that any two $r_i$ and $r_j$ are statistically independent. Then, the shortest possible algorithm $R_k$ that outputs a description of $\mathcal{T}$ when fed $\mathcal{T}$ will be of the same lenght as $\sharp{R_k}$. In other words, random time-series are algorithmically incompressible. This is one way to define randomness. Algorithmic complexity also allows us to define \emph{complexity} in a succinct way: if a description of $\mathcal{T}$ that is shorter in length than $\mathcal{T}$ itself cannot be obtained by $R_k$ in polynomial time (polynomial in the length of $\mathcal{T}$), then we say that $\mathcal{T}$ is \emph{complex}. Wolfram (2002) has conjectured that all Class III and Class IV ECAs are NP-complete\footnote{ Class III and IV ECAs are NP-complete in the following sense: suppose one inputs some initial string into a given ECA and lets the ECA 'run' upto $n$-steps. The string at the $n$-th step will not, in polynomial time, tell us what string we inputed in the beginning.}.
\par
On the other hand, say that $R_k$ is inputing a time-series that describes the positions of atoms in a crystal. Because of geometric regularity of a crystal, $\sharp(R_k)$ is going to be quite small. To see this explicitly, we consider a time-series
$$\mathcal{T} = \{-1,0,1,-1,0,1, -1,0,1, \ldots\}$$
Here there is only one block of data $\{-1,0,1\}$ and $\mathcal{T}$ is constructed by "repeating" this block indefinitely. Therefore, $R_k$ will input this and output something that would say "regular". 
In the Wolfram classifcation, Class I and Class II ECAs are algorithmically compressible while Class III and Class IV are very dramatically not\footnote{During \emph{NKS 2003: Conference and Minicourse}, Wolfram demonstrated the ineffectiveness of standard compression softwares to compress Rule 30, an example of Class III ECA. The output was longer than the input!}. We see that algorithmic complexity allows us to answer the question we raised in the Section 2.1 in a limited way-- since $R_k$ is assumed to be performed on a universal Turing machine, and all universal computers are equivalent to each other, one way to classify ECAs would be to look at the compressibility of any ECA with respect to a given compression algorithm. 
\par
Wolfram goes on to suggest something much stronger about these issues; He calls it the \emph{Principle of Computational Irreducibility} (PCI). PCI asserts that if the length of $R_k$ is the same as the description of $\mathcal{T}$, then dynamical evolution of $O_i$ is, itself, a computation of the same length. He also conjectures that all Class III and Class IV ECAs are universal computers (universal Turing machines). Both of these taken together imply that almost all phenomena that is "structured" but not 'simple' are equivalent to universal computers!
\section{The action chain}
In our discussion so far, we have not really talked about the \emph{physics} of the measurement process-- a meter is a physical object that interacts with the observable in order to 'spew-out' a measurement. In classical physics, the meter is only an inference device and the observable is given an \emph{a priori} ontology. Crutchfield terms this one-way flow of information from the observable to the meter as the \emph{Einstein flow}. On the other hand, in a quantum physical situation, the observable does not exist (ontologically) unless there is a meter that performs a measurement- by that very act, it drastically changes the inference capacity of the observer (\emph{the uncertainty principle}) and sometimes the very ontology of the observable is changed (\emph{the Scrodinger cat paradox}). This two-way exchange of information between an observable and a meter is termed by Crutchfield as the \emph{Heisenberg flow}. This also, famously, gives us a definition of a \emph{meter as something capable of inducing a dynamics on the observable that is being measured}. I denote this action of the meter on the observable by the action chain:
$$
\begin{CD}
O_i @<\textrm{acts}<< M_j
\end{CD}
$$
\par
Now, as Crutchfield argues, this situation is not exclusively seen at the quantum level. In classical dynamical systems that are sensitively dependent on the choice of intial conditions, meters become capable of inducing dynamics on the observable in such a way that the time-series thus obtained "reflects" both the intrinsic dynamics of the observable and the "extrinsic" dynamics that are imposed by the meter. Crutchfied (1994) illustrates the essentially entangled nature of the system-under-study and the environment in which the system is placed with a nice example:
\begin{quote}
Consider the gravitational effect of an electron at the ``edge'' of the known Universe (17 billion light years) on a terrestrial game of billiards. Assume, for simplicity, that during a given shot the game is energy conserving over half an hour and that the balls are hit sufficiently hard to cause a few collisions per second. The unpredictability of the billiards' state can be conservatively estimated as an information loss rate of approximately 1 bit per second. The unncertainty caused by the existence or nonexistence of the electron at the edge of the universe leads to total unpredictability in about six minutes.
\end{quote}
We reformulate this example in terms of meters. Suppose we are interested in recording the initial conditions of a game of billiards. The observable in question is the position of an arbitrary ball on the table. We do this using a meter that emits a photon, hits that ball, and comes back to us, thereby enabling us to detect the position of the ball. Now the photon will impart a small additional momentum on that ball. This additional momentum will change the initial condition that we are intent on observing slightly. However, since a billiard ball is a dynamical system that is extremely sensitive to the choice of initial conditions, this tiny change in momentum would lead to a very different time-evolution of the position of the ball, very different from what the system would have evolved to, had the measurement not been made.
\par
Let us look into such systems further. Suppose we have an observable whose dynamical behavior is dependent on the performance of a measurement. The main question is the commensurability of 
$$
\begin{CD}
O_i @<\textrm{acts}<< M_j
\end{CD}
$$
and
$$
\begin{CD}
O_i @>\textrm{input}>> M_j.
\end{CD}
$$
In other words, $M_j$ detects information about the dynamics of $O_i$. Yet, at the same time, it contributes to the dynamics itself. Therefore, is there a way by which we could construct a recognition device that captures this entaglement of the inference and action chain? Certainly, algorithmic complexity does not help us here very much- all that $R_k$ does (when viewed as an algorithm) is sort $O_i$ into one of the bins, depending upon the length of $R_k$; it says nothing about the origin of the dynamics that a given observable displays. It seems that this problem, as formulated, is intractable. In Rosen's language (Rosen, 2001a, 2001b), this entanglement of the action and inference chain is an example of \emph{impredicativity} that distinguishes complex from simple systems.
\par
By an impredicative (complex)system, Rosen means a system that is essentially bootstrapped to the environment/meter/observer-- recall that the only way we get to construct a time-series (from which one can reconstruct the geometry of the phase space of the observable) is by successive application of the meter to the observable. However, this time-series is only partially faithful to the original, nascent dynamics of the observable; it also incorporates the dynamics induced by the measurement. Crutchfield argues (as seen from his billiard-and-electron example) that this very impredicativity is what disallows us to effectively predict the behavior of most nonlinear complex system. 
\par
In this Chapter, we have provided a very condensed review of the (inter-linked) notions of complexity, computation and measurement and the notion of inference in a context/meter dependent environment. We have seen that algorithmic complexity theory is not likely to help us ``disentagle'' the contributions of the dynamics induced by the meter on an observable from the nascent dynamics of the observable itself. We have seen however that algorithmic complexity theory is likely to help us in providing a universal invariant charateristic of complex systems by which we can classify them. In summary, this charateristic is the length of the algorithm that gives us a description of a system that is complete when "fed" a time-series that is inferred from the observable's dynamics. 

\chapter{Scale, Measurement and the System/Subsystem Distinction}
\section{Epistemology vs. ontology}
From the dicussion in the preceeding Chapter, it is clear that \emph{any} theory of coordination in a sufficiently complex system must address the problem of measurement very seriously. We have seen in Chapter 2 the essentially incommensurate nature of measurement when viewed as an inference chain in contradistinction to when viewed as an action chain. A measurement, paradoxically, is capable of limiting our inferential ability and our capacity to predict. This is equally valid in quantum mechanical systems and classical mechanical systems that are sensitively dependent on the initial data supplied to them.
\par
There is another aspect to the problem of measurement that seriously challenges orthodoxy-- the system/subsystem distinction, an understated reductionist assumption, becomes unassailable because of the very nature of the act of measurement. In this Chapter, we discuss this problem.\footnote{The discussion that follows is set through more conventional notions of ontology versus episetmology. However, it might be better phrased in the terminology of context-dependence versus context independence. Context independence is essentially an ontological assumption. Context dependence, on the other hand, is subsuming this hypothesis in favor of the assumption that all knowledge about the world is essentially knowledge-through-some-observer. The context independent perspective argues that our knowledge about the world is about the things-that-are. However, when measurement is viewed as action (the action chain of Chapter 2), all knowledge-about-the-world is infused and entangled with the very act of measurement. Therefore, the context-independent perspective is considerably weakened.} \par
Like complexity, the very definition of the words \emph{system} and \emph{subsystem} are hard to come by. That is what \emph{is} a system, and what do we mean by its \emph{subsystems}? I wish to argue that this very question is ill-phrased; from a non-reductionistic viewpoint the distinction is purely epistemological. The germs of the arguments presented here can be found in Rosen (1978, 2001a, 2001b) and Turvey (2002) even though Rosen's resolution of the issue is somewhat different from what we would conclude. (Rosen in \cite{R1} resolves the system/subsystem distinction into \emph{state-restricted}, \emph{observable-restricted} and \emph{dynamics-restricted} subsystems of a system.)
\section{System/susbsystem dichotomy and self-reflexivity}
Consider a collection of non-elemental (non-analytic in the Hegelian sense) units of any arbitrary sort. The constraint on this assemblage is that the elements in this collection are "linked" via some physical interaction.(A simple example of such a collection is a box of Helium atoms interacting with each other via collisions and intramolecular forces.) Recall that a meter as defined in Chapter 2 "detects" a certain class of observables via an instaneous measurement at and in a certain spacetime scale. In a sense, meters and observables have to have a certain "affinity" for each other for the observer to obtain non-trivial measurements\footnote{For example, one could stick a Geiger counter in a tub of pure water. One is likely to get a set of $0$ in $\mathcal{T}$! We could operationalize the definition of affinity between meters and observables in the following way: we say that a meter and an observable have an affinity for each other if and only if we obtain a set of non-zero entries in $\mathcal{T}$.}.
\par
If the meter is made to detect a certain class of observables via an (instantaneous) measurement at (and in)a certain spacetime scale, then the system is what is specified by that class of observables. One could then impose another meter on the assemblage that, say, detects another class of observables at a larger/longer spacetime scale. The system, then, is what is specified by that set of observables, and the previous set of observables specify the (new) system's subsystem(s). By indefinitely repeating this procedure, one obtains an indefinitely large set of classes of observables, each class of observables specifying a system. The notion of a subsystem is, then, derived from the ordering of this set of classes of observables according to the spacetime scales that the meters (which specify and are specified by the system) operate on.
\par
Of course, what is being alluded to in the argument above is the notion of \emph{context dependence}. Depending upon the context (the space-time scale at which the meter is operating), the definitions of what a system and its subsystems are changes. This context-dependence of the system/subsystem distinction is, then, a pseudodichotomy and not ontological, because the distinction is a matter of an observer invoking a meter at a certain spacetime scale which then limits our capacity of inference maximally (Chapter 2). If this is the case, then in sufficiently complex systems, one should not look for a linear causal chain between the system and its subsystems. The notion of causality is a valid notion only in the cases where things that are being said to be causally linked have an ontological status.
\par
This poses two challenges, one philosophical and the other methodological:
\begin{itemize}
\item{\emph{The notion of explanation}. Traditionally, the explanation of a phenomenon in the sciences is taken to be the explication of causality between an arbitrary number of scales of observation of a given phenomenon, where one scale is seen as "uniquely privileged" in terms of being the residence of analytic units. If the notion of causality between different scales, in this case between a system and its subsystems, does not hold true any longer, then we have to replace our notion of explanation accordingly. (Baas \& Emmeche (1997) suggest a new notion of explanation for complex systems that is context-dependent.)}
\item{\emph{The pursuit of natural laws} Having said that one cannot seek an explanation of a phenomenon in the traditional sense of a linear causal chain emanating from a privileged scale and tying the successive (space/time) ordered scales together, any enterprise of scientific enquiry that is grounded in realism is committed to the belief in an essential \emph{lawfulness} of the observed phenomena at \emph{all} scales (Kugler \& Turvey, 1987). Each set of observables specifying a system at a given scale is to be lawfully constrained by the set of observables at other scales that specify other systems that could be subsystems of the given system. This lawfulness should not be construed to suggest any causal connection between the different scales.}
\end{itemize}

\chapter{Free Parameters and ``Tight Models''}
\
In the last Chapter, we noted that a theory of multiscale coordination that is not reductionistic must be based on lawful relations that cut across scales and yet resist the temptation of ascribing a causal link between the different scales. What we were alluding to was a search for regularities upon which empirical science is based, and yet a regularity that cannot be ascribed to causal relations in a heirarchical setting of scales. 
\par
Recall that in Chapter 1, we proposed two touchstones for such a theory. One of the touchstones was that a good theory of multiscale coordination will be such that in that theory, the \emph{deux ex machina} of free parameter fine tuning will be minimal; a popular motto among theorist and modellers is that ``with enough parameters one can fit an elephant.'' So, how are we to avoid ``explaining'' everything with a maximally large set of free parameters? After all, with enough free parameters that are liable to be fine-tuned by hand, we could build an extremely cumbersome model that ``inputs'' a lot of information (i.e. allows for a lot of fine-tuning of free parameters by hand) and ``outputs'' limited information about the system that is being modelled. In order to set a tone for a discussion of these issues, we fix a definition of a \emph{free parameter}. We define a free parameter\footnote{synonomously, \emph{control parameters} or \emph{model parameters}} in a dynamical system as a non-state variable whose values cannnot be deduced from the model itself, and is, instead, ``put in by hand'' \emph{a priori} way.
\par
For example, consider the HKB potential
\begin{equation}
V(\phi) = a\cos \phi - b\cos 2\phi.
\end{equation}
Clearly, the values of $a$ and $b$ drive the dynamical layout (decides the attractors/repellors) but the value of $a$ or $b$ cannot be deduced from the dynamics itself. The analogy of free parameters, as Rosen (1978) points out, is with genotypes in biology. Genotypes specify phenotypes (analogous to the state variables) but not the other way round.
\par
It is astonishing how much of modelling in the natural sciences boil down to choosing and fine-tuning free parameters by hand. For example, the Standard Model of particle physics works only when 17 free parameters (masses of quarks, mixing angles, Planck's constant, \ldots) are fine-tuned by hand, i.e., they are specified by the theorists rather than specified by the theory (Smolin, 1995). This has often been cited as the single-biggest failure of particle physics. 
\par
In physical biology, Kugler, et.al. (1980) following a proposal by Bernstein proposed a set of criteria that any viable theory of coordination should satisfy,; one of the criterion was having a minimally small set of free parameters. Inspite of this, all models of coordination that have been proposed so far (mostly based on coupled oscillator models) have not been able to effectively stop the proliferation of free parameters. In fact, in the Beek-Peper-Daffertshoffer two-tiered model, the number of free parameters actually increase due to the addition of the "neural substrate" dynamics to the classical effector level dynamics.
\par
In a sense, the problem of having a minimally small set of free parameters is essentially a problem of finding an \emph{entailment structure for free parameters}. By that, we mean the following: Let $F = \{\alpha_1, \alpha_2, \ldots, \alpha_k\}$ be a set of $k$ free parameters: $\alpha_i = f (\alpha_j)$ for arbitrary $i,j$ and $1 \leq i \leq k$ and $1 \leq j \leq k$. (Masses obeying an allometric realtion in a  dynamical model are obvious examples of members of a set of entailed free parameters.) In other words, an entailment structure for free parameters refer to a scenario where the values of any arbitrary free parameter in the model depend upon the value of any other free parameter in the same model thereby reducing the number of free parameters that should be finetuned by hand to one.
\par
One way to provide an entailment structure to the set of free parameters that has been extensively discussed in the control theory literature is by making free parameters functions of state variables (cf. Isacc, 1957). The problem with this approach is the following: free-parameters drive the state-variables and determine the dynamical layout. If we let the dynamical layout specify the value of free-parameters, we arrive at very constrained and bootstrapped dynamics that is likely to end up with a logical circularity making the problem intractable.

\chapter{Formal Similitude and Dynamical Similarity }
In Chapter 2, 3 and 4, we have identified various problems that need to be addressed by any candidate theory of multiscale coordination: the role of a measurement device and its ramifications in defining a complex system, the essentially self-reflexive nature of the system/subsystem distinction and lastly, the problem of finding entailment structures for free parameters in order to stop the proliferation of the same. In this Chapter, I identify one possible candidate that seems to adhere to the two touchstones that were set in Chapter 1. This candidate theory is based on the notion of formal similitude, which I, following Rosen (1988), call the \emph{principle of dynamical similarity}.
\section{Allometry}
Biological arguments based on similarity and congruence of different physiological scales are not new. Principally, the work of d'Arcy Thompson, (2001, reprint) and von Holst, (1973, reprint) has shown us how to utilize notions of structural invariance (that different layers of physiological organization have similar structures) and morphological invariance (that there are a few primary 'morphs', and that all observed morphologies are results of coordinate transformations of these primary morphs). In essence, in both cases, the key notion has been that of invariance of structure and form. In more modern guises, these ideas have taken the name of ``fractal organization'' of physiological structure. They reflect the truism that all scales of neural organization exploit mechanisms that are structurally embedded in each other and nested together.Typically, systems that exhibit morphological invariance and structural invariance also exhibit the phenomena of \emph{allometry}.
\par
Allometry is defined as the following: suppose one has two observables $x$ and $y$. If $y = px^q$, where $p,q$ are rational numbers, then we say that the observables are \emph{allometrically scaled}. For Rosen, allometrically scaled relations (in recent terminology, \emph{power laws}), between state variables were examples of \begin{quote}
the ultimate synergy, \ldots, [that is] manifested by totally constrained systems
\end{quote}
Recall the example of the Standard Model of particle physics. As noted before, there are 17 free parameters in the model. However, as Smolin (1995) notes (based on the work of Dirac, Tipler and Carter), most observables at the scale of astrophysical and cosmological processes such as 'typical' stellar masses obey allometric relations with respect to those free parameters, even though there is no strict mechanistic theory that explains why this should be the case. (Dirac termed these allometric relations the ``law of large numbers'')
\section{Formal similitude}
Invariance, in another guise, has been a fundamental principle in modern physics for a long time-- recall that Einstein's special and general relativity are statements about the invariance of the fundamental equations of physics under certain (covariant coordinate) transformations. However, the principle of invariance in modern physics has been the invariance of the (mathematical) forms of the laws of physics rather than invariance of actual physical structure, as in d'Arcy Thompson or von Holst. I term such principles of invariance \emph{formal similitude}. 
\par
In the context of biological similitude, Rosen (1978), was one of the first people to point out the usefulness of formal similitude. He illustrated this "principle" in the following way: within a given cell, the concentration of chemical species $x_i$  can be described by a mass-action rate law 
\begin{equation}
\dot{x_i} = f_i(x_1, \ldots, x_n) 
\end{equation}
where $f_i$ are determined by specific reactions between the chemical species. Now $\{x_1, x_2, \ldots, x_n\}$ are the basic observables of cell chemistry. Rosen noted that a map $\rho: \{x_1, x_2, \ldots, x_n\} \to \mathbb{R}$ is, then, also an observable. Let $F = F(x_i,\ldots, x_n)$ and $G = G(x_1,\ldots, x_n)$ be two such observables.
We can write
\begin{equation}
\dot{F} = \Phi(F, G)  \textrm{ and } \dot{G} = \Psi(F, G)
\end{equation}
for "appropriately chosen functions" $\Phi$ and $\Psi$. Notice that \emph{(5.2) has the same form as the mass-action rate laws (5.1); they are mass-action rate laws for the new observables $F$ and $G$}. For Rosen, this indicates a "universality" of the form of the equations that observables satisfy. 
\par
In the context of multiscale coordination, in a sense what we are after is \emph{one} canonical form of dynamical system that observables at all scales should satisfy. This idea of formal similarity is made precise by the notion of \emph{topological conjugacy}. 
\section{Principle of dynamical similarity}
Topological conjugacy is defined as follow (Smale, 1967): Let $f: A\to A$ and $g: B\to B$ be two maps. $f$ and $g$ are said to be \emph{topologically conjugate} if there exits a homeomorphism $h: A\to B$ such that $f \circ h = h \circ g$.
In other words, the following diagram commutes:
\[
\begin{CD}
A               @>g>>         A\\
@VhVV                         @VVhV\\
B               @>f>>         B
\end{CD}
\]
The hypothesis that is being advanced here is the following: Let
\begin{eqnarray}
\ldots \nonumber\\
g_{i-1}: S_{i-1} \to S_{i-1}, g_{i-1} = {\dot x}_{i-1}\nonumber\\
g_i: S_i \to S_i, g_i =  {\dot x}_i\nonumber\\
g_{i+1}: S_{i+1} \to S_{i+1}, g_{i+1 }= {\dot x}_{i+1} \nonumber\\
\ldots 
\end{eqnarray}
be equations of motion at different scales indexed by $\{\ldots, i-1, i, i+1, \ldots\}$ and $\{ \ldots, S_{i-1}, S_i, S_{i+1}, \ldots \}$ be observables at scales indexed by $\{\ldots, i-1, i, i+1, \ldots\}$. Then there exists the set of conjugacy homeomorphisms $\{ \ldots, h_{i-1}, h_{i}, h_{i+1}, \ldots \}$ across scales such that the following diagram commutes:
\[
\begin{CD}
\cdots  S_{i-1}  @>h_{i-1}>>   S_i   @>h_{i}>>  S_{i+1}  \cdots        \\  
        @Vg_{i-1}VV            @Vg_{i}VV        @Vg_{i+1}VV           \\
\cdots  S_{i-1}  @>h_{i-1}>>   S_{i} @>h_{i}>>  S_{i+1}  \cdots
\end{CD}
\]
In other words, the following set of relations hold true:
\begin{eqnarray}
\ldots \nonumber\\
g_{i} \circ h_{i-1} = h_{i-1} \circ g_{i-1} \nonumber \\
g_{i+1} \circ h_i = h_i \circ g_{i} \nonumber \\
\ldots
\end{eqnarray}
This is the principle of dynamical similarity. The germ of this formulation of the principle is to be found in Rosen (1978, 1988). 
\section{Predictions from dynamical similarity}
The principle of dynamical similarity, in summary, is about homeomorphisms between observables. Now we investigate the commensurability of this principle with the inference chain. Recall from Chapter 3 that a meter $M_j$ acts on an observable $S_i$ as an inference device in the following way:
$$
\begin{CD}
(S_i @>\textrm{input}>> M_j)^t @>\textrm{output}>> \{r_1, r_2, \ldots, r_t\} (:=\mathcal{T}).
\end{CD}
$$
Intuitively, a commensurate relation between this inference chain and the principle of dynamical similarity suggests that time series obtained by $M_j$ from $S_i$ is going to be similar to the time series obtained by another meter $M_{j+1}$ acting on the observable $S_{i+1}$. This suggests the first testable prediction of a theory of multiscale coordination based on the principle of dynamical similarity: \emph{Time series obtained from the dynamics of different strata of physiological organization are predicted to be similar to each other}.\footnote{This intuitive notion can be formalized by the notion of a (topological) duality between the principle of dynamical similarity and the inference chain. One can treat the principle of dynamical similarity in (5.4) and the action chain as \emph{duals}.}
\par
One way that this could be experimentally checked in a laboratory situation is to let subjects perform a given task while the experimenter monitors the performance of that task as well as monitors other physiological variables that are relevant to the performance of the given task.
This method can be summarized by the following algorithm:
\begin{itemize}
\item{Take a multiscale system that is capable of utilyzing different layers of orgainzation in the performance of a given task. Obtain time-series at different scales while that task is being performed} Using techniques of phase space reconstruction, one can obtain the geometries from the time-series (Packard et.al., 1980)
\item{Model the reconstrutced flows in the phase spaces thus obtained with low dimensional equations of motion for each scale.}
\item{Examine if there is a suitable transformation by which those ``reconstructed equations of motions'' into one another.}
\end{itemize}
We also see that with the principle of dynamical similarity, one is able to lawfully constrain the values of free parameters and provide an entailment structure for the entire set of free paramters. 
\par
In order to see how the principle of dynamical similarity constraint the behavior of free parameters, let us look into an example taken from Strogatz (1994/2000). The logistic map (a discrete version of the Verhulst equation) is given by
\begin{equation}
x_{n+1} = rx_n(1 - x_n).
\end{equation} The free parameter in the logistic map is $r$.
The claim that we want to prove is that the logistic map is topologically conjugate to another biologically relevant discrete difference equation, the quadratic map, 
\begin{equation}
y_{n+1} = y_n^2 +c,
\end{equation}
where $c$ is the free parameter. These maps are topological conjugate because there exists the homeomorphism $x_n = -\frac{1}{r} y_n + \frac{1}{2}$.
\par
We immediately see that if the quadratic map and the logistic map are topologically conjugate (they are), then the free parameters $r$ and 
$c$ are related as
\begin{equation}
c = \frac{r(2 - r)}{4}.
\end{equation}
Therefore, the principle of dynamical similarity entails the two free parameters in a lawful way.
\par
There is another auxillary testable consequence of the principle of dynamical similarity. Recall that the principle states, in essence, that the dynamics of the $i$-th scale "mimics" the dynamics of the $i+1$-th scale. This implies that if one scale has a stable limit cycle dynamic, then \emph{all} other scales will inherit the same dynamical layout. On the other hand, if one scale shows chaotic behavior, then all other scales will have chaotic dynamics.
\par
From the perspective of multiscale coordination in movement, this has important implications. Certain studies of the human rhythmic movement have identified regimes of chaotic behavior experimentally. For example, studies of COP fluctuations show a highly self-similar fractal behavior with non-trivial Hurst exponents (Riley et.al., 1997). However, the origin of such dynamics is unclear. If the principle of dynamical similarity holds true, then we know that dynamics of scales ``smaller'' or ``larger'' than that of COP fluctuations will exhibit chaotic dynamics as well. The challenge for the experimentalist is to identify these scales and demonstrate that the underlying dynamics have precisely the same behavior as that of the COP fluctuation scale. On the other hand, if certain tasks at an effector level are seen to have a stable dynamical behavior, then all other scales are likely to have the same "structural stability'' (Smale, 1967). 
\chapter{Conclusion}
In this Project, we examined a number of cognate issues that, taken together, pertain to the creation of a non-reductionistic theory of multiscale coordination. Being primarily an investigation into the philosophy of theoretical biology and physics, I have not attempted to present any particular model of a given phenomenom. I have, instead, attempted to sketch, in broad strokes, what a viable model should look like and the particulars it must address.
\par
I began in Chapter 1 by addressing the two main reasons why one should look for models of multiscale coordination: the observed stochasticity of motor control and Wilson's principle of evolutionary holism. We presented two touchstones that a theory of nonreductionistic multiscale coordination must satisfy: (1) the role of measurement in such systems must be more than marginal and the system/subsystem distinction has to be addressed as an epistemological rather than an ontological problem; (2) the number of free parameters must be minimally small but maximally entailed.
\par
Chapter 2 addressed the issue of complexity of a given system in terms of a new synthesis of the work of Rosen (impredicative complexity) on one hand and the work of Wolfram, Chaitin and Crutchfield (computational complexity) on the other hand. We introduced a novel notational system that enabled us to view a meter both as an inference device as well as a machine capable of introducing dynamics on the very observable that is being measured. Thereby, we show that the so-called inference chain and the action chain are essentially incommensurate. This leads to the immediate consequence that the system/subsystem distinction is not at all rigid; the measurement process, when viewed as an action chain, interferes with the very object we are trying to infer about as well as demonstrates the context-dependency of the notion of a complex system.
\par
Chapter 3 examines the role of scale in measurement and demonstrates that the system/subsystem distinction is essentially self-reflexive. Thereby, we showed that a "system" and its "subsystems" are epistemological constructs. The purpose of this argument was to challenge the orthodoxy of reductionism and models that are built on the understated assumption of "priveleged scales". We argued that the self-reflexive nature of the system/subsystem distinction poses serious challenges to the notion of explanation and forces us to account for lawful behavior in alternative ways. 
\par
Chapter 4 was an examination of the role of free parameters in a given model--  we argued the nessecity of obtaining entailment structures for free parameters that limit the proliferation of free parameters and thereby increase the effectiveness of a model. 
\par
Chapter 5 was the culmination of the Project. I presented one candidate for the theory of multiscale coordination that obeyed the principle of dynamical similarity. We showed that such a theory makes two predictions each of which is falsifiable in a laboratory situation.

\newpage
\begin{Huge}
\flushleft{\textbf{Bibliography}}
\end{Huge}
\vspace{20mm}\\Alexander, D.M., and Globus, G.G. (1996). Edge-of-chaos dynamics in recursively organized neural systems. In E.M. Cormac and M.I. Stamenov (Eds.) \emph{Fractals of Brain, Fractals of Mind: In Search of a Symmetry Bond} pp.31--73. Amsterdam/Philadelphia : John Benjamins.\\
\\
d'Arcy Thompson, W. (2001). \emph{On Growth and Form}. Reprint. Oxford: Oxford University Press.\\
\\
Baas, N. \& Emmeche, C. (1997). On emergence and explanation. \emph{Intellectica}, \textbf{2} (25), 67--83. Also available as Santa Fe Institute Working Paper 97-02-008.\\
\\
Barrow, J.D. and Tipler, F.J. (1986). \emph{The Anthropic Cosmological Principle}. Oxford: Oxford University Press.\\
\\
Beek, P.J., Peper, C.E., \& Daffertshofer, A. (2002). Modeling rhythmic interlimb coordination: Beyond the Haken-Kelso-Bunz model. \emph{Brain and Cognition},\textbf{48}, 149--165.\\
\\
Brooks, R.A. (1991). Intelligence without representation. \emph{Artificial Intelligence}, \textbf{47}, 139--159.\\
\\
Chaitin, G.  (1987). \emph{Algorithmic Information Theory}. Cambridge: Cambridge University Press.\\
\\
Clark, A. (1999). \emph{Being There}. Cambridge, Mass.: MIT Press.\\
\\
Crutchfield, J.P. (1994). Observing Complexity and The Complexity of Observation. In H. Atmanspacher (Ed.) \emph{Inside versus Outside,: Workshop on Endo/Exo Physics}, pp. 234 -- 272, Berlin: Springer-Verlag.\\
\\
Dirac, P.A.M. (1937). The cosmological constants. \emph{Nature}, \textbf{139}, 323--326.\\
\\
Dirac, P.A.M. (1938). A new basis for cosmology. \emph{Proceedings of the Royal Society}, \textbf{A165}, 199--217.\\
\\
Frank, T.D., Daffershoffer, A., \& Beek, P.J. (2000). Multivariate Ornstein-Uhlenbeck processes with mean-field dependent coefficients: Applications to postural sway. \emph{Physical Review E}, \textbf{63}, 011905-1--011905-16.\\
\\
Frank, T.D., Daffertshofer, A., Peper, C.E., Beek, P.J., \& Haken, H. (2000). Towards a comprehensive theory of brain activity: Coupled oscillator systems under external forces. \emph{Physica D}, \textbf{144}, 62--86.\\
\\
Gelfand, I.M., Gurfinkel, V.S., Tsetlin, M.L. \& Shik, M.L. (1971). Some problems in the analysis of movements. In I.M. Gelfand, V.S. Gurfinkel, S.V. Fomin \& M.L. Tsetlin (Eds.) \emph{Models of Structural-Functional Organization of Certain Biological Systems} pp. 329 -- 345. Cambridge, MA: MIT Press.\\
\\
Haken, H., Kelso, J.A.S., \& Bunz, H. (1985). A theoretical model of phase transitions in human hand movements. \emph{Biological Cybernetics}, \textbf{51}, 347--356.\\
\\
von Holst, E. (1973). \emph{The Behavioral Physiology of Animal and Man}. Coral Gabes, FL: University of Miami Press.\\
\\
Isaacs, R. (1963).\emph{Differential Games With Applications to Problems of Warfare}. New York: Dover.\\
\\
Kay, B.A., Saltzman, E.L., \& Kelso, J.A.S. (1991). Steady-state and perturbed rhythmical movements: A dynamical analysis. \emph{Journal of Experimental Psychology: Human Perception and Performance}, \textbf{17}, 183--197.\\
\\
Kay, B.A., Kelso, J.A.S., Saltzman, E.L., \& Sch\"{o}ner, G. (1987). Space-time behavior of single and bimanual rhythmical movements: Data and limit cycle model. \emph{Journal of Experimental Psychology: Human Perception and Performance}, \textbf{13}, 178--192.\\
\\
Keele, S.W. (1968). Movement control in skilled motor performance. \emph{Psychological Bulletin}, \textbf{70}, 387 -- 403.\\
\\
Kugler, P.N., Kelso, J.A.S. \& Turvey, M.T. (1980). On the concept of coordinative structures as dissipative structures: I. Theoretical lines of convergence. In G.E. Stelmach \& J. Requin (Eds.), \emph{Tutorials in Motor Behavior} pp. 3--47. New York: John Holland.\\
\\
Kugler, P.N. \& Turvey, M.T. (1987). \emph{Information, Natural Law and the Self-Assembly of Rhythmic Movement}. Hillsdale: Lawrence-Erlbaum.\\
\\
Mineev-Weinstein, M. B. (1996). Noise, fractal growth, and exact integrability in nonequilibrium pattern formation. In M. Millonas (Ed.), \emph{Fluctuations and Order: The New Synthesis} pp. 239--257. New York: Springer.\\
\\
Mitra, S., Riley, M.A. \& Turvey, M.T. (1997). Chaos in human rhythmic movement. \emph{Journal of Motor Behavior}, \textbf{29}, 195 -- 198.\\
\\
Packard, N.H., Crutchfield, J.P., Farmer, J.D. \& Shaw, R.S (1980). Geometry from a time-series. \emph{Physical Review Letters}, \textbf{45}, 712 --717.\\
\\
Riley, M.A., \& Turvey, M.T. (2001). Variabilty and determinism in motor behavior.\emph{Journal of Motor Behavior}, to appear.\\
\\
Riley, M.A., Wong, S., Mitra, S., \& Turvey, M.T. (1997). Common effects of touch and vision on postural parameters. \emph{Experimental Brain Research}, \textbf{117}, 165--170.\\
\\
Rosen, R. (1978). \emph{Fundamentals of Measurement and Representation of Natural Systems}. New York: North-Holland.\\
\\
Rosen, R. (1988). Similarity and dissimilarity: A partial overview. \emph{Human Movement Science}, \textbf{7}, 131--153.\\
\\
Rosen, R. (2001a). \emph{Essays on Life Itself}. New York: Columbia University Press.\\
\\
Rosen, R. (2001b). \emph{Life Itself}. New York: Columbia University Press.\\
\\
Sch\"{o}ner, G., Haken, H., \& Kelso, J.A.S. (1986). A stochastic theory of phase transitions in human hand movement. \emph{Biological Cybernetics}, \textbf{53}, 247--357.\\
\\
Scott, A. (1995). \emph{Stairway to the Mind: The Controversial New science of Consciousness}. New York: Springer.\\
\\
Smale, S. (1967). Differentiable dynamical systems. \emph{Bulletin of the American Mathematical Society}, \textbf{73}, 747--817.\\
\\
Sol\'{e}, R. \& Goodwin, B.C. (2000). \emph{Signs of Life: How Complexity Pervades Biology}.New York: Basic Books.\\
\\
Smolin, L. (1995). \emph{Cosmology as a problem in critical phenomena}. Electronic preprint, arXiv:gr-qc/9505022.\\
\\
Strogatz, S. H. (1994/2000). \emph{ Nonlinear Dynamics and Chaos: With Applications to Physics, Biology, Chemistry and Engineering}. Cambridge, Mass.: Perseus.\\
\\
Tomita, K. (1984). Coarse graining revisited- The case of macroscopic chaos. In Y. Kuramoto (Ed.) \emph{Chaos and Statistical Methods} pp.2--13. Berlin/Heidelberg: Springer.\\
\\
Tuffillano, N.B., Reilly, J. \& Abbott, T. (1992). \emph{An Experimental Approach to Nonlinear Dynamics and Chaos}. Mass.: Addison-Wesley.\\
\\
Turvey, M.T. (1990). Coordination. \emph{American Psychologist}, \textbf{45}, 938--953.\\
\\
Turvey, M.T. (2002a). \emph{Lectures on Perception and Action}. Unpublished manuscript.\\
\\
Turvey, M.T. (2002b). Personal communication.\\
\\
Turvey, M.T. \& Shaw, R.E. (1995). Ecological foundations of cognition: I. Symmetry and specificity of animal-environment systems. \emph{Journal of Consciousness Studies}, \textbf{6}, 95--110.\\
\\
Wolfram, S. (1994). \emph{Cellular Automata and Complexity: Collected Papers}. Cambridge, MA: Perseus Publishing.\\
\\
Wolfram, S. (2002). \emph{New Kind of Science}. Urbana-Champagne, IL: Wolfram Media.

\newpage
\begin{center}
\textbf{Acknowledgements}
\end{center}
I would like to thank my University Scholar Advisory Committee, Professors Michael Thomas Turvey (Department of Psychology), Stuart Jay Sidney (Department of Mathematics) and Juha Javanienen (Department of Physics), for guidance and inspiring a sense of scholarship. Hopefully, this Project bears the imprints of their patient training. I gratefully acknowledge their permission to draw extensively from the material prepared by me for two Research in Psychology (PSYC 304) courses during  Spring 2002 semester (\emph{Dynamics of Neural Nets} supervised by Prof. Whitney Tabor) and Summer 2003 (\emph{Coordinative Structures and Complexity} supervised by Prof. Michael Thomas Turvey). 
\par
I also wish to thank members of the UConn Center for Ecological Study of Perception and Action and Haskins Laboratories, New Haven, CT for wonderful discussions; in particular, Dr. Bruno Galantucci and Prof. Robert Shaw shared their insights with me and pressed the neccesity of investigating the philosophy and methodology of physiological psychology, Prof. Whitney Tabor lent his expertise in attempts to understand the philosophy behind the neural networks approach to modelling the brain and behavior and Mr. Theo Rhodes convinced me that ``somebody has to get to the bottom of murky waters that complexity theory is.''
\par
 This Project was substantially supported by the United States National Science Foundation grant SBR 97- 28970 (Principal Invesitigator: M.T. Turvey) and the University of Connecticut Honors Program.
\par
This Project, like all of my other academic endeavors, is dedicated to my father. Bapi convinced me, very early on in my life, that being a scholar is sublimely rewarding. 
\par
\emph{Postscript added on 27th February 2004}: I wish to thank Annie Olmstead (UConn) for her patient help with proof-reading this version.

\end{document}